\documentstyle[epsf,floats,eqsecnum,twocolumn,prl,aps]{revtex}
\newcommand{\bS}{\mbox{\boldmath $S$}}
\newcommand{\bn}{\mbox{\boldmath $n$}}
\newcommand{\bm}{\mbox{\boldmath $m$}}
\newcommand{\bl}{\mbox{\boldmath $l$}}

\newcommand{\stimes}{\!\times\!}
\newcommand{\bff}{\mbox{\boldmath $f$}}
\newcommand{\bg}{\mbox{\boldmath $g$}}
\begin{document}
\draft
\preprint{}
\title{Alternating spin chains with singlet ground states}
\author{ Takahiro Fukui\cite{Email}} 
\address{Institute of Advanced Energy, Kyoto University,
Uji, Kyoto 611, Japan}
\author{Norio Kawakami}
\address{Department of Applied Physics,
Osaka University, Suita, Osaka 565, Japan} 
\date{April 10, 1997: Revised June 23, 1997}
\maketitle
\begin{abstract}
We investigate low-energy properties of the alternating spin 
chain model composed of spin $s_1$ and $s_2$ with a singlet ground
state.  After examining the spin-wave spectrum in detail, we 
map low-energy spin excitations to the O(3) non-linear sigma model
in order to take into account quantum fluctuations. Analyzing 
the topological term in the resulting sigma model, we discuss how the 
massless or massive excitations are developed, especially according
to the topological nature of the alternating spin system.
\end{abstract}
\pacs{PACS: 75.10.Jm, 05.30.-d, 03.65.Sq} 
\section{Introduction}
One-dimensional quantum spin systems show various interesting
phenomena, having been continually providing hot topics.
The most important example is the spin one-half
antiferromagnetic Heisenberg chain,
which was exactly solved by the Bethe-ansatz method.
It is now well known that this model shows critical behavior
specified by the level-1 Wess-Zumino-Witten model.
Keeping the integrability by adding multiple terms, 
higher-spin  models were constructed\cite{TakBab} and 
their critical behavior was also classified.
Progress in a slightly different direction
was made by Haldane\cite{Hal,HalRevAff,HalRevFra,HalRevTsv}, 
who conjectured that 
general higher-spin models behave in quite distinct ways 
according to whether the spin of the models is integer 
or half-odd integer. The massive phase of integer 
spin models, so-called 
the Haldane phase, has been indeed observed by experiments,
and now is understood as the valence-bond-solid state\cite{VBS}.
The problem whether a given quantum spin system is massless 
or massive provides nontrivial and interesting issues, 
which have been stimulating the study of
quantum phase transitions.

More recently, the mixed spin chains with 
an alternating array of two kind of
spins have attracted much current interest\cite{Exp,Alt}.
The systems experimentally found so far 
\cite{Exp} show a magnetically ordered
ground state even for the antiferromagnetic case.
For such ordered systems, quantum fluctuations 
may not play a crucial role. However,
it is quite interesting to address the question what happens 
when the magnetic order disappears due to frustrations, etc.  
If such systems are realized experimentally, 
quantum fluctuations should certainly play  
an important role, exhibiting interesting phenomena such as
the competition between  the massless and massive phases. 
These systems may thus provide  
a new interesting paradigm of quantum spin chains.  
In this connection, an exactly solvable alternating spin chain,
which has a singlet ground state,  
has been studied in detail\cite{AltExa}.

The alternating spin models can also be regarded as 
a high concentration limit of magnetic impurities in
spin chains.
In this respect, the problem of the alternating spin models  
is closely related to that for the  spin chains with impurities,
which have been extensively studied for e.g. 
the two-leg ladder system\cite{RICE,NTAFT,FNST},
the spin-Peierls system \cite{RRDR,FTS}, etc.
In fact by changing the concentration of periodic array of
impurities, we can naturally interpolate the alternating spin systems
and spin systems with dilute impurities, and discuss 
their characteristic properties in a unified way 
\cite{FukKaw}. 

In this paper, we study low-energy properties of the 
mixed spin chain models with an alternating array of
two kind of spins. We deal with two kind of
alternating spin models with a singlet ground
state.  By introducing quantum fluctuations
 to the spin-wave excitations, we 
investigate  low-energy properties of the system
by resorting to O(3) non-linear sigma model techniques.
We then discuss how the 
massless or massive excitations are developed in the 
alternating spin models
by analyzing the topological term in the sigma model.

The paper is organized as follows.
In Section II, two types of the models for alternating spin chains
are introduced,  and some rigorous statements are
given for the excited states as well as the ground state.
In Section III, we calculate the spin-wave dispersion relation, 
and in Section IV, by converting the model
to the non-linear sigma model, we study the 
low-energy characteristics of the spin excitations. 
The last section is devoted to summary of the 
paper.

A brief report on preliminary results for this paper 
is given in \cite{FukKaw}, where the effects of massive modes were not
taken into account in the nonlinear sigma model approach.

\section{Models and Basic Properties}
In this section, we first introduce two kind of 
alternating spin models with a singlet
ground state, and then study them by means of Lieb-Schultz-Mattis
(LSM) theorem.

The models we study in 
this paper are  the quantum spin models with nearest neighbor
interaction in one dimension. 
 The corresponding Hamiltonian is defined by
\begin{equation}
H=\sum_{j=1}^{N_a}J_j\bS_j\cdot\bS_{j+1},
\label{Ham}
\end{equation}
where $N_a$ is the number of lattice sites,
and the spin at the $j$-th site is denoted by $s_j$.
We will investigate the following two types of 
alternating spin-chain models by setting
\begin{eqnarray}
&&\hbox{\underline{Model A:}}
\nonumber\\
&&s_j=\left\{
\begin{array}{ll} 
s_1 & \hbox{ for } j=1 \\
s_2 & \hbox{ for } j=2,3 
\end{array}
\right.
\hbox{ (mod 3) },
\nonumber\\
&&J_j=J,
\nonumber\\
&&N_a=6N,
\label{ParA}\\
&&\hbox{\underline{Model B:}}
\nonumber\\
&&s_j=\left\{
\begin{array}{ll} 
s_1 & \hbox{ for } j=1,2 \\
s_2 & \hbox{ for } j=3,4 
\end{array}
\right.
\hbox{ (mod 4) }, 
\nonumber\\
&&J_j=\left\{
\begin{array}{ll} 
J(1-\gamma_1) & \hbox{ for } j=1 \\
J[1+(\gamma_1+\gamma_2)/2] & \hbox{ for } j=2,4\\
J(1-\gamma_2) & \hbox{ for } j=3 \\
\end{array}
\right.
\hbox{ (mod 4)},
\nonumber\\
&&N_a=4N.
\label{ParB}
\end{eqnarray}
In both cases, the spin chain is composed by two kind
of spins $s_1$ and $s_2$, which can take 
either integer or half integer spin. Note that we also include the 
effect of the bond-alternation for the Model B, which 
was originally introduced by Tonegawa et al.
\cite{TONE} In what follows, 
we assume the antiferromagnetic couplings $J>0$ and
$0<\gamma_1,\gamma_2<1$.

Let us start by specifying the  ground state properties.
Since spins are on a bipartite lattice,
we can apply the Marshall theorem\cite{LieMat,AffLie,Aue,FukKaw} 
to (\ref{Ham}), that is, we prove that
the ground state is singlet without degeneracy for both models.
Next consider the properties of excitations above the ground state
by applying the LSM theorem\cite{LSM}. 
Note that the Hamiltonian is invariant under 
3- (4-)site translation for the Model A (B).
Therefore, together with the fact that the ground state is unique, 
we can construct an excited state of $O(1/N)$ when 
$s_{\rm uc}=$ odd-integer,
where $s_{\rm uc}$ is the summation of the spins in the unit cell
\begin{equation}
s_{\rm uc}=\left\{\begin{array}{ll}
s_1+2s_2&\hbox{ for A}\\
2(s_1+s_2)&\hbox{ for B}
\end{array}\right.
\label{Suc}
\end{equation}
According to the above formula, we can state\cite{FukKaw} that
the Model A can be gapless if $s_1$ is a half-integer.
In other cases where $s_1$ is an integer for the Model A,
or $s_1$ and $s_2$ are arbitrary for the Model B,
we cannot claim anything by the LSM theorem, which suggests 
that such systems may be gapful, as is the case for 
the ordinary Haldane system.

\section{Spin-wave spectrum}

So far we have seen that the Model A can show different phases, 
according to whether $s_1$ is integer or half-integer.
However, the LSM theorem itself cannot distinguish a gapless 
phase from a gapful phase with degenerate ground states.
In order to clarify this point, 
we wish to use a complementary approach based on  effective field 
theory.  For this purpose, 
we first investigate the dispersion
relation of the spin-wave spectrum for the model, 
which allows us to judge whether the system can be 
mapped to the non-linear sigma model.
The next subsection is devoted to deriving the spin-wave Hamiltonians
by the Holstein-Primakoff mapping, and they are diagonalized in the
subsection B. We then show some examples of the 
spin-wave spectrum calculated numerically.

\subsection{Spin-wave Hamiltonian}
\subsubsection{Model A}

For the usual uniform spin chain ($s_1=s_2$ case for the present
model),
it is sufficient to introduce the two kind of bosons, corresponding to
bipartite lattices.
However, for the model with $s_1\ne s_2$ we have to 
introduce the six kind of bosons such that
\begin{eqnarray}
&&S^z_{6j+m}=(-)^m\left[
n_{6j+m}^{(m)}-s_{m}\right],
\nonumber\\
&&S^+_{6j+m}=\left\{
\begin{array}{ll}
\sqrt{2s_{m}-n_{6j+m}^{(m)}}a_{6j+m}^{(m)}&
\hbox{ for odd }m\\
a_{6j+m}^{(m)\dagger}\sqrt{2s_{m}-n_{6j+m}^{(m)}}&
\hbox{ for even }m\end{array}\right.,
\label{HolPri}
\end{eqnarray}
where $j=0,1,\cdots,N-1$, $m=1,2,\cdots,6$,
$s_m=s_1~(s_2)$ for $m=1,4$ (otherwise) and
$n_{6j+m}^{(m)}=a_{6j+m}^{(m)\dagger}a_{6j+m}^{(m)}$.
In what follows, we approximate the boson Hamiltonian up to 
quadratic order in order to derive the dispersion relation
of spin wave excitations.
In this assumption, the resultant Hamiltonian can be diagonalized 
in the momentum space.  
For this purpose, let us introduce the Fourier-transformed 
operators
\begin{equation}
a_{6j+m}^{(m)}=\frac{1}{\sqrt{N}}
\sum_{k=0}^{N-1}\exp\left[
\frac{2\pi i}{N}\frac{k}{6}(6j+m)\right]a_{k}^{(m)},
\label{BosFou}
\end{equation}
with $N$ being defined in eq.(\ref{ParA}).
Substituting these equations, we find
\begin{eqnarray}
H_{\rm sw}=\sum_{k=0}^{N-1}\sum_{l,m=1}^6&&
(a_l^\dagger h_{lm}a_{m}+a_lh_{lm}a_{m}^\dagger
\nonumber\\
&&\qquad+a_l^\dagger\Delta_{lm}a_{m}^\dagger+a_l\bar{\Delta}_{lm}a_{m}),
\label{SpiWavHam}
\end{eqnarray}
where we have used a notation,
\begin{equation}
a_m=\left\{
\begin{array}{ll}
a_{k}^{(m)} &\hbox{ for odd $m$}\\
a_{-k}^{(m)}& \hbox{ for even $m$}
\end{array}\right.
\label{aSho}
\end{equation}
for short, and $6\times6$ matrices $h$ and $\Delta$ are given by
\begin{eqnarray}
&&h=Js_2\hbox{diag}\left(
1,\frac{1+\alpha}{2},\frac{1+\alpha}{2},
1,\frac{1+\alpha}{2},\frac{1+\alpha}{2}\right),
\nonumber\\
&&\Delta=\frac{Js_2}{2}\left(
\begin{array}{cccccc}
0&\sqrt{\alpha}\rho&&&&\sqrt{\alpha}\bar{\rho}\\
\sqrt{\alpha}\rho&0&\bar{\rho}&&&\\
&\bar{\rho}&0&\sqrt{\alpha}\rho&&\\
&&\sqrt{\alpha}\rho&0&\sqrt{\alpha}\bar{\rho}&\\
&&&\sqrt{\alpha}\bar{\rho}&0&\rho\\
\sqrt{\alpha}\bar{\rho}&&&&\rho&0     
\end{array}\right) .
\label{hDelA}
\end{eqnarray}
We have defined in these equations 
\begin{equation}
\alpha=\frac{s_1}{s_2},\quad \rho=e^{ip}
\label{AlpGam}
\end{equation}
with $p=(2\pi/N_a)k$.
Matrix elements not written explicitly in eq.(\ref{hDelA}) are all 0.
Note that the operators (\ref{aSho}) as well as the matrix 
$\Delta$ depend on 
the momentum, though we have not explicitly denoted it.
\subsubsection{Model B}

In a  similar way to the Model A, we now 
write down the spin-wave Hamiltonian for the Model B.
Introduce the Holstein-Primakoff mapping with the use of 
four kinds of bosons similar to eq.(\ref{HolPri}) by replacing 
$6\rightarrow4$.
With the same Fourier-transformed operators as eq.(\ref{BosFou}),
we have the spin-wave Hamiltonian similar to eq.(\ref{SpiWavHam})
for which $h$ and $\Delta$ are $4\times4$ matrices defined by
\begin{eqnarray}
&&h=\frac{Js_2}{2}\hbox{diag}\left(
\alpha\Gamma_1+\Gamma,\alpha\Gamma_1+\Gamma,
\alpha\Gamma+\Gamma_2,\alpha\Gamma+\Gamma_2
\right),\nonumber\\
&&\Delta=\frac{Js_2}{2}\left(
\begin{array}{cccc}
0&\alpha\Gamma_1\rho&0&\sqrt{\alpha}\Gamma\bar{\rho}\\
\alpha\Gamma_1\rho&0&\sqrt{\alpha}\Gamma\bar{\rho}&0\\
0&\sqrt{\alpha}\Gamma\bar{\rho}&0&\Gamma_2\rho\\
\sqrt{\alpha}\Gamma\bar{\rho}&0&\Gamma_2\rho&0
\end{array}\right) ,
\label{hDelB}
\end{eqnarray}
where
$\Gamma_i=1-\gamma_i$ and
$\Gamma=1+(\gamma_1+\gamma_2)/2$.

\subsection{Diagonalization}
In the previous subsection, we have derived the spin-wave Hamiltonians
up to the quadratic order in boson operators.
We can diagonalize these Hamiltonians by the Bogoliubov
transformation.

Before discussing the spectrum of the present models,
let us recall the alternating spin-chains with ferrimagnetic
ground state\cite{Alt}, i.e, the same model given by 
(\ref{Ham}), but with
\begin{equation}
s_{2j-1}=s_1,\quad s_{2j}=s_2,\quad J_j=J. 
\end{equation}
The spin-wave dispersion relation for this model is 
\begin{equation}
\omega_\pm=\pm|s_1-s_2|+
\left[(s_1-s_2)^2+4s_1s_2\sin^2p\right]^{1/2}.
\end{equation}
Thus the excitation spectrum 
is given by  a quadratic function for small momentum.
It is only the case for 
$s_1=s_2$ that it has a linear dispersion.
\begin{figure}[h]
\epsfxsize=7cm 
\centerline{\epsfbox{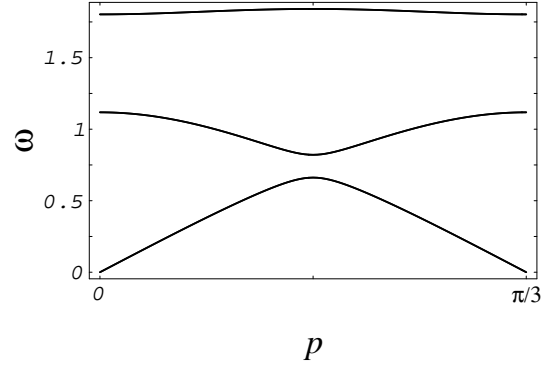}} 
\vspace{0.5cm}
\caption{Spin-wave spectrum of the Model A in unit $J$ 
as functions of the momentum $p$ 
for $s_1=1/2$ and $s_2=1$.}
\label{f:ModA}
\end{figure}
\begin{figure}[h]
\epsfxsize=7cm 
\centerline{\epsfbox{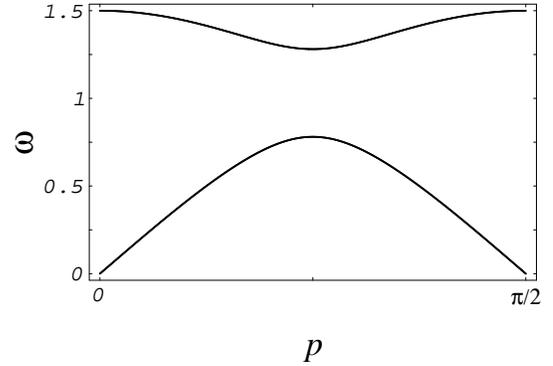}} 
\vspace{0.5cm}
\caption{Spin-wave  spectrum of the Model B in unit $J$
as functions of the momentum $p$ 
for $s_1=1/2$ and $s_2=1$.
Bond-alternation parameters are $\gamma_1=\gamma_2=0$.}
\label{f:ModB}
\end{figure}

Now turn to the present model.
We show the spectrum obtained  
numerically in Figs.\ref{f:ModA} and \ref{f:ModB}.
We can see that the dispersion relation $\omega_0\propto\sin p$ 
defined in $p\in[0,\pi]$ for the uniform chain 
is folded 3 (2) times for the Model A (B), 
and interactions among such modes produce gaps at the crossing 
point. 
Though we here presented only the case with $s_1=1/2$ and $s_2=1$,
the qualitative features are the same for other cases $s_1\ne s_2$.
Contrary to this, in the case $s_1=s_2$, interactions between 
the different modes
disappear and the single dispersion $\omega_0$ appears.
What is remarkable is that the lowest mode has a linear dispersion, 
in contrast to the ferrimagnetic case, without an excitation gap. 
\begin{figure}[h]
\epsfxsize=7cm 
\centerline{\epsfbox{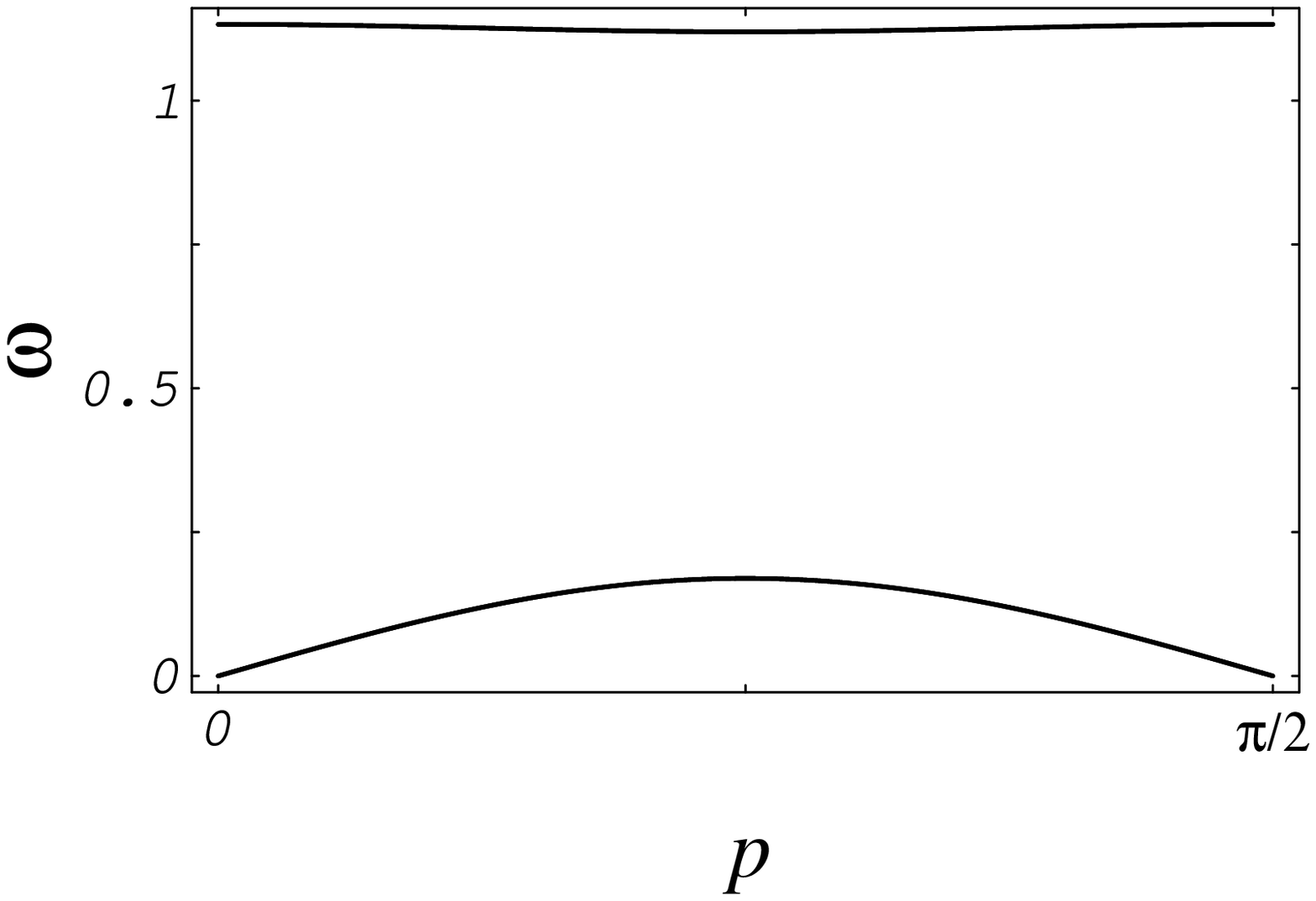}} 
\vspace{0.5cm}
\caption{The same figure as Fig.\ref{f:ModB}, but with 
$\gamma_1=\gamma_2=0.9$.}
\label{f:ModBbondP}
\end{figure}
\begin{figure}[h]
\epsfxsize=7cm 
\centerline{\epsfbox{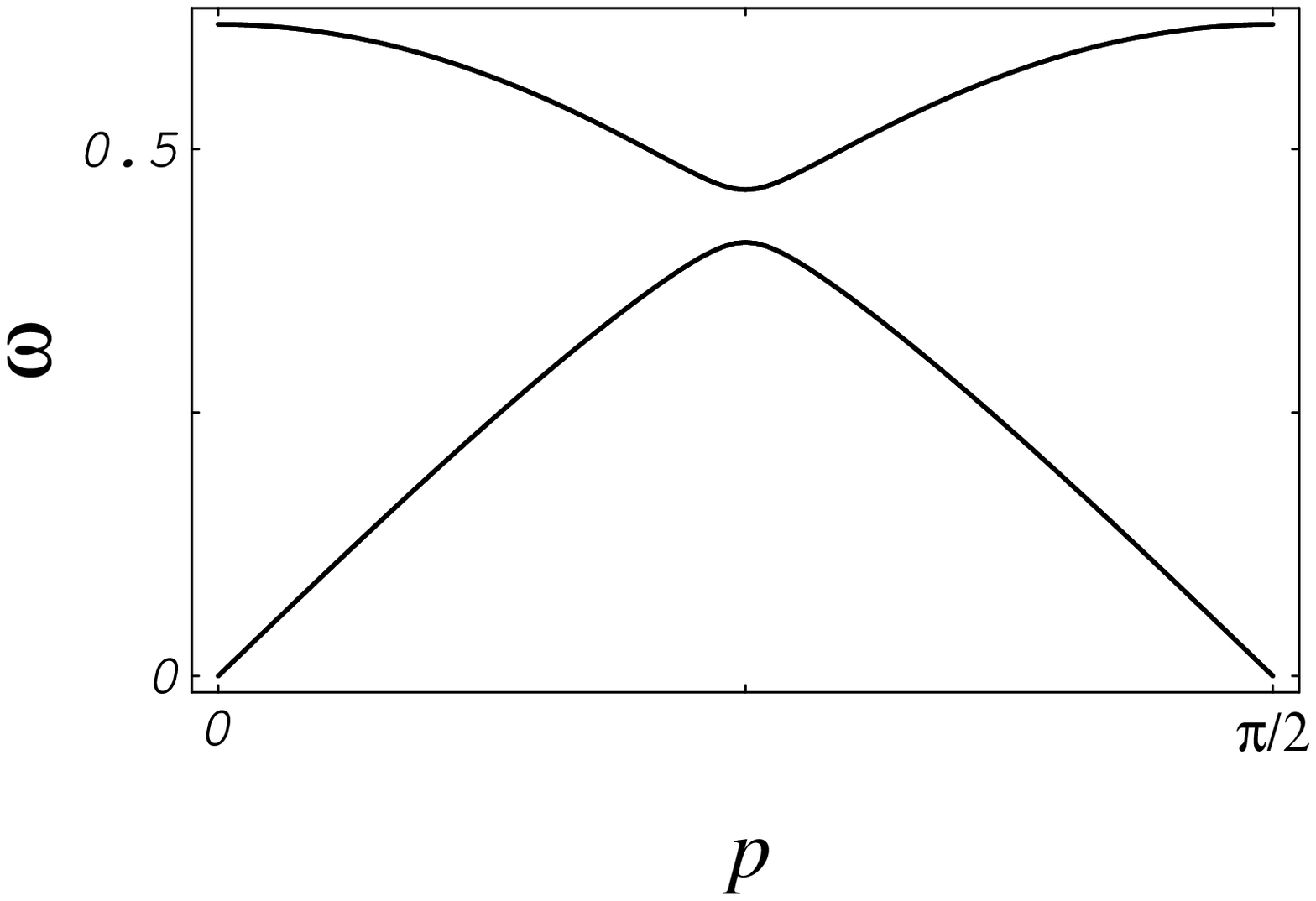}} 
\vspace{0.5cm}
\caption{The same figure as Fig.\ref{f:ModB}, but with 
$\gamma_1=\gamma_2=-0.9$.}
\label{f:ModBbondM}
\end{figure}
This property persists even if we include the bond-alternation,
as can be seen in Figs.\ref{f:ModBbondP} and \ref{f:ModBbondM}. 
However, 
it goes without saying that because we have such a spin-wave spectrum, 
we cannot conclude a gapless spectrum of the system
via the lowest-order spin-wave analysis.
This is due to the fact that
higher order quantum corrections play a crucial role whether
or not the system is indeed gapless.
What we would like to emphasize in this section is that 
the lowest mode has a linear spectrum, for which 
we are ensured to use non-linear sigma model techniques
in order to investigate the characteristic properties of the models.

\section{Non-linear sigma model approach}

As was shown in the previous section, the spin excitations
in both models A and B are characterized by the linear spectrum in 
the low energy regime.
This suggests that we can map these models to O(3) non-linear sigma
model (NLSM) in order to incorporate 
the quantum effects in a leading order
approximation. Although the NLSM has been successfully applied to the
usual uniform spin chain\cite{Hal,HalRevAff,HalRevFra,HalRevTsv}, 
there appear in the present models
some other massive modes in addition to the lowest
massless spin excitations,
 making the problem somewhat complicated.
Therefore, it is not straightforward how to 
incorporate the 
 effects of massive modes into the massless mode to
obtain the effective NLSM.
In this section, we propose a way to treat such effects,
following the method recently developed in ref.\cite{DEMPR} for
spin-ladder systems.

The starting action is 
\begin{equation}
S=S_B+S_H,
\end{equation}
where
\begin{eqnarray}
&&S_B=-i\sum_{j=1}^{N_a}(-)^{j+1}s_jw[\bn(j)],
\nonumber\\
&&S_H=\frac{1}{2}\sum_{j=1}^{N_a}J_js_js_{j+1}
\int\!\!d\tau\left[\bn(j+1)-\bn(j)\right]^2.
\end{eqnarray}
In the following, we would like to treat the two models separately.
\subsection{Model A}

For the Model A, the spin-wave spectrum has a three-band structure 
 when $s_1\ne s_2$.
Therefore, we introduce the sigma-model fields as,
\begin{equation}
\bn(3j+a)=\bm(3j+a)+(-)^{3j+a+1}a_0\bl_a(3j+a),
\label{MasModA}
\end{equation}
where $a_0$ is a lattice constant, $j=0,1,\cdots,N-1$ and $a=1,2,3$.
Note that whether the present models are
massive or massless is  essentially determined by the coefficient 
of the topological term, and may not be sensitive to 
how many fields we consider in the effective theory. 
However, it is expected that the introduction of 
three kind of fields $l_1,l_2$ and $l_3$ improves our low-energy 
effective theory, by incorporating the effects due to 
the massive modes to some extent:
e.g. a better approximation may be  
obtained for the spin-wave velocity, as is seen below.  
In what follows we assume that the several fields thus introduced
 are sufficiently smooth
functions so that we can take the continuum limit of them.

First, let us derive the expression for the Berry phase term
in the continuum limit, 
\begin{eqnarray}
S_B=&&~\sum_{a=1}^3\sum_{j=0}^{N-1}(-)^{a+1}is_a
\nonumber\\
&&\quad\times\left\{\omega[\bn(6j+a+3)]-\omega[\bn(6j+a)]\right\}
\nonumber\\
=&&\sum_{a=1}^3\sum_{j=0}^{N-1}(-)^ais_a
\nonumber\\
&&\times\int\!\!d\tau\delta\bn(6j+a)\cdot
[\bn(6j+a)\stimes\partial_\tau\bn(6j+a)],\nonumber\\
\end{eqnarray}
where $s_a=s_1,s_2,s_2$ for $a=1,2,3$, respectively, and 
$\delta\bn(6j+a)\equiv\bn(6j+a+3)-\bn(6j+a)$. 
We replace the difference by the differentiation,
$\delta\bn(6j+a)\sim3a_0
\partial_x\bm(6j+a)+(-)^a2a_0\bl_a(6j+a)+O(a_0^2)$,
where $a_0$ is the lattice constant.
Substituting this formula, we then have
\begin{eqnarray}
S_B=&&\frac{is}{2}
\int\!\!d^2x\bm\cdot(\partial_x\bm\stimes\partial_\tau\bm)
\nonumber\\
&&\qquad
+is_2\sum_{a}\int\!\!d^2xf_a\bl_a\cdot(\bm\stimes\partial_\tau\bm),
\label{ConLimBer}
\end{eqnarray}
where $s=s_1$ and
\begin{equation}
\bff=\frac{1}{3}\left(\begin{array}{c}
\alpha\\1\\1\end{array}\right),
\end{equation}
with $\alpha$ defined in eq.(\ref{AlpGam}).

Let us next take the continuum limit of interaction terms,
\begin{eqnarray}
S_H=&&\frac{J}{2}\sum_{a=1}^3\sum_{j=0}^{2N-1}s_as_{a+1}
\nonumber\\
&&\qquad\times\int\!\!d\tau\left[\bn(3j+a+1)-\bn(3j+a)\right]^2.
\end{eqnarray}
Here, note that 
$\bn(3j+a+1)-\bn(3j+a)\sim a_0\partial_x\bm(3j+a)
+a_0(-)^{3j+a}
\left[\bl_{a+1}(3j+a)+\bl_a(3j+a)\right]+O(a_0^2)$.
Substituting this expression and neglecting oscillating terms,
we have
\begin{eqnarray}
S_H=&&\frac{Ja_0s_2^2}{2}\int\!\!d^2x
\bigl[K(\partial_x\bm)^2
\nonumber\\&&\qquad\qquad
+\sum_ag_a\bl_a\cdot\partial_x\bm+\sum_{a,b}\bl_aL_{ab}\bl_b\bigr],
\label{ConLimInt}
\end{eqnarray}
where $K=(2\alpha+1)/3$, $\bg^t=(0,0,0)$ and 
\begin{equation}
L=\frac{1}{3}\left(
\begin{array}{ccc}
2\alpha&\alpha&\alpha\\
\alpha&1+\alpha&1\\
\alpha&1&1+\alpha
\end{array}
\right).
\label{MatL}
\end{equation}
Here we have introduced $\bg$ for later convenience (Model B).

Integrating the field $\bl$, we end up with the effective Lagrangian 
density for $\bm$,
\begin{eqnarray}
{\cal L}=&&\frac{1}{2g}\left[
v(\partial_1\bm)^2+\frac{1}{v}(\partial_2\bm)^2\right]
\nonumber\\&&\qquad\qquad
+\frac{\theta}{8\pi}\epsilon_{\mu\nu}\bm
\cdot(\partial_\mu\bm\stimes\partial_\nu\bm),
\label{Lag}
\end{eqnarray}
where
\begin{eqnarray}
&&\theta=2\pi is\left(1+\frac{s_2}{s}\bg^tL^{-1}\bff\right),
\nonumber\\
&&g=\frac{1}{s_2}
\frac{1}{\sqrt{
\left(K-\frac{1}{4}\bg^tL^{-1}\bg\right)\bff^tL^{-1}\bff}},
\nonumber\\
&&v=Ja_0s_2\sqrt{\frac{K-\frac{1}{4}\bg^tL^{-1}\bg}{\bff^tL^{-1}\bff}},
\label{GenTGV}
\end{eqnarray}
provided that $L$ is a symmetric matrix.
For the Model A, by the use of the following inverse matrix of $L$ 
defined in eq.(\ref{MatL})
\begin{equation}
L^{-1}=
\frac{3}{4\alpha}\left(
\begin{array}{ccc}
\alpha+2&-\alpha&-\alpha\\
-\alpha&\alpha+2&\alpha-2\\
-\alpha&\alpha-2&\alpha+2
\end{array}\right),
\end{equation}
we have the following formula,
\begin{eqnarray}
&&\theta=2\pi is_1 ,
\nonumber\\
&&g=\frac{2}{s_2}\frac{3}{\sqrt{(2\alpha+1)(\alpha^2-2\alpha+4)}},
\nonumber\\
&&v=2Ja_0s_2\sqrt{\frac{2\alpha+1}{\alpha^2-2\alpha+4}}.
\label{TheGVA}
\end{eqnarray}
It should be noted that in the case of $s_1=s_2\equiv s$, the above 
formulae reduce to  those of the 
uniform spin-$s$ spin chain, $\theta=2\pi is$, $g=2/s$ 
and $v=2Ja_0s$.

It is interesting to compare the formula (\ref{TheGVA}) as 
those in Appendix A, which was previously obtained by assuming 
a single $l$-field (i.e., neglecting the effect of
 massive modes)\cite{FukKaw}.
The coefficient of the topological term is the same,
$\theta=\widetilde\theta$, which makes the above formula reliable.
The difference of the bulk quantities appears in the factors depending
on $\alpha$. 
\begin{table}[h] 
\begin{center}
\renewcommand{\arraystretch}{1.2}
\begin{tabular}{ll|lllll}
$\hfill s_1$& & 1/2 & 1 & 3/2 & 2 & 5/2\\ 
\hline
         &$v_{\rm sw}$  & 1.000 & 1.061 & 0.878 & 0.707 & 0.582 \\
$s_2=1/2$&$v$           & 1     & 1.113 & 1     & 0.866 & 0.761 \\
         &$\widetilde v$& 1     & 1.25  & 1.4   & 1.5   & 1.571 \\
\hline
         &$v_{\rm sw}$  & 1.488 & 2.000 & 2.179 & 2.121 & 1.952 \\
$s_2=1  $&$v$           & 1.569 & 2     & 2.219 & 2.236 & 2.138 \\
         &$\widetilde v$& 1.6   & 2     & 2.286 & 2.5   & 2.667 \\
\end{tabular}
\end{center}
\caption{ Spin-wave velocity for the Model A.
The velocity denoted by $v_{\rm sw}$ is calculated by the
spin-wave dispersion relation $\omega \sim v_{\rm sw}p$
for small $p$. It is compared with $v$ given by 
eq.(\ref{TheGVA}) in sec. IV. For reference, 
we also show $\widetilde v$ given by eq.(\ref{OldTGVA})
for which the effects of massive modes are  neglected. }
\label{t:VelA}
\end{table}
   From Table \ref{t:VelA}, 
we found $v$ closer to the value $v_{\rm sw}$.
We can thus see that the approximation is 
improved by the introduction 
of three $l$-fields rather than a single $l$-field.

Based on the above observations,  we now 
arrive at the following conclusion for the Model A. The 
topological term is 
controlled only by the spin $s_1$, and this characteristic
property is independent of the approximations we have used.
Namely, if $s_1=$ integer (half-integer), the model is to be 
massive (massless). This is indeed consistent with the result by
Lieb-Shultz-Mattis theorem in sec.II.
\subsection{Model B}

In the spin-wave analysis, we have 
found a two-band structure in the spectrum for the Model B, 
and so we here introduce the fields, 
\begin{equation}
\bn(j)=\bm(j)+(-)^{j+1}a_0\stimes\left\{
\begin{array}{c}\bl_1(j)\\\bl_2(j)\end{array}\hbox{ for } 
\begin{array}{c}j=1,2\\j=3,4\end{array}\hbox{ (mod 4)}\right..  
\label{MasModB}
\end{equation}
Manipulations for passing to the continuum limit are 
performed in parallel to the Model A.
First, note that the continuum limit of the Berry phase terms  
takes the same form as eq.(\ref{ConLimBer}), but with
\begin{equation}
s=\frac{s_1+s_2}{2}, \quad
\bff=\frac{1}{2}\left(\begin{array}{c}\alpha\\1\end{array}\right).
\end{equation}
Here, subscript $a$ in eq.(\ref{ConLimBer})
runs from 1 to 2, because we have two modes now.
Next, interaction terms are also calculated in the same form as
eq.(\ref{ConLimInt}) for which the corresponding 
parameters are given as
\begin{eqnarray}
&&K=\frac{1}{4}\left(
\Gamma_1\alpha^2+2\Gamma\alpha+\Gamma_2\right),
\nonumber\\
&&\bg=\left(
\begin{array}{c}
\Gamma\alpha-\Gamma_1\alpha^2\\
\Gamma\alpha-\Gamma_2
\end{array}\right),
\nonumber\\
&&L=\frac{1}{2}\left(\begin{array}{cc}
2\Gamma_1\alpha^2+\Gamma\alpha&
\Gamma\alpha\\
\Gamma\alpha&
2\Gamma_2+\Gamma\alpha
\end{array}\right),
\end{eqnarray}
where $\Gamma_i\equiv1-\gamma_i$ and 
$\Gamma\equiv 1+(\gamma_1+\gamma_2)/2$.

It is straightforward to derive the explicit formulae 
for general $\gamma_1$ and $\gamma_2$,
by eq.(\ref{GenTGV}). However, their appearance is rather complicated
generally, so that we here write down them for the simple 
case $\gamma_1=\gamma_2\equiv\gamma$. We thus have
\begin{eqnarray}
&&\theta=\pi i(s_1+s_2)
\frac{4(1+\gamma)s_1s_2}
{(s_1+s_2)^2+\gamma(s_1-s_2)^2 },
\nonumber\\
&&g=\frac{1}{s_1s_2}
\frac{(s_1+s_2)^2+\gamma(s_1-s_2)^2}
{\sqrt{(1+\gamma)[(s_1+s_2)^2+\gamma(s_1^2-6s_1s_2+s_2^2)]}},
\nonumber\\
&&v=4Ja_0s_1s_2
\sqrt{\frac{(1-\gamma)(1-\gamma^2)}
{(s_1+s_2)^2+\gamma(s_1^2-6s_1s_2+s_2^2)}}.
\nonumber\\
\label{TheGVB}
\end{eqnarray}
Now we wish to discuss the above formulae in 
several limiting cases.
First, by setting $s_1=s_2=s$, they are reduced 
 to $\theta=2\pi is(1+\gamma)$,
$g=2/s\times(1-\gamma^2)^{-1/2}$ and $v=2Ja_0s(1-\gamma^2)^{1/2}$,
i.e, the usual formula for the uniform chain with bond-alternation.
Next set $\gamma=0$ for arbitrary $s_1$ and $s_2$.
Then we have
\begin{eqnarray}
&&\theta=4\pi i\frac{s_1s_2}{s_1+s_2},
\nonumber\\
&&g=\frac{s_1+s_2}{s_1s_2},
\nonumber\\
&&v=4Ja_0\frac{s_1s_2}{s_1+s_2}, \quad{\rm for}~\gamma=0,
\label{TGBzer}
\end{eqnarray}
We predict from this expression that {\it at $\gamma=0$ the system is
massive in general}.
For example, $\theta=4\pi i/3$ in the case $s_1=1/2$ and $s_2=1$.
Note that the bulk quantities $g$ and $v$ are
different from $\widetilde g$ and $\widetilde v$, respectively,
defined by eq.(\ref{OldTGVB})
which were previously derived\cite{FukKaw}
by neglecting massive modes from  the beginning. On the other hand
the coefficient of the topological term is the same in 
both cases, $\theta=\widetilde\theta$.
\begin{table}[h] 
\begin{center}
\renewcommand{\arraystretch}{1.2}
\begin{tabular}{ll|lllll}
$\hfill s_1$& & 1/2 & 1 & 3/2 & 2 & 5/2\\ 
\hline
         &$v_{\rm sw}$  & 1.000 & 1.333 & 1.500 & 1.600 & 1.666 \\
$s_2=1/2$&$v$           & 1     & 1.333 & 1.5   & 1.6   & 1.667 \\
         &$\widetilde v$& 1     & 1.491 & 1.936 & 2.332 & 2.687 \\
\hline
         &$v_{\rm sw}$  & 1.333 & 2.000 & 2.400 & 2.666 & 2.857 \\
$s_2=1$  &$v$           & 1.333 & 2     & 2.4   & 2.667 & 2.857 \\
         &$\widetilde v$& 1.491 & 2     & 2.500 & 2.981 & 3.440 \\
\end{tabular}
\end{center}
\caption{Spin-wave velocity for the Model B. Parameters are same 
as Table \ref{t:VelA}, but with $v$ given by 
eq.(\ref{TheGVB}), 
and $\widetilde v$ given by eq.(\ref{OldTGVB}). }
\label{t:VelB}
\end{table}
By observing that $v$ reproduces $v_{\rm sw}$ exactly in 
Table \ref{t:VelB}, we naturally expect 
that the assumption (\ref{MasModB}) in our 
approach works rather well, making the above formula more reliable
than the previous treatment with a single-$l$ field approximation.

Next, let us discuss  the effects of the bond-alternation.
First, we recall that for the usual uniform spin-$s$ chain (see
the above formula for $s_1=s_2$ case), 
$\theta=0\rightarrow4\pi is$ if $\gamma$
varies $\gamma=-1\rightarrow1$.
Namely, the model realizes $2s$ 
massless points during this process\cite{AffHal}.
However, it is difficult in general to predict the exact value of
the bond-alternation parameter $\gamma$ which brings about
massless phases. For example, when $s=1$, it is predicted 
by Affleck and Haldane\cite{AffHal} that a
massless phase occurs at $\gamma=1/2$, while 
the numerical calculation suggests a different
value $\gamma=0.25$\cite{KatTan}.
This is because for systems with explicitly broken 
translational symmetry, there are no strong restrictions by this 
symmetry and the coefficient of the topological term can
take arbitrary values. This makes such values less reliable.
Nevertheless it is believed that the number of massless points 
predicted by sigma model techniques  itself should be correct, 
which has been indeed realized for the $s=1$ case 
by numerical calculations \cite{KatTan}.

Keeping this in mind, we would like to
observe how the topological term $\theta$ behaves
for for $s_1 \neq s_2$ case. From 
eq.(\ref{TheGVB}), we can see that $\theta$ is an increasing
function of $\gamma$, 
\begin{equation}
\theta=0\rightarrow \pi i(s_1+s_2)\frac{4s_1s_2}{s_1^2+s_2^2},
\end{equation}
according to $\gamma=-1\rightarrow1$.
\begin{figure}[h]
\epsfxsize=7cm 
\centerline{\epsfbox{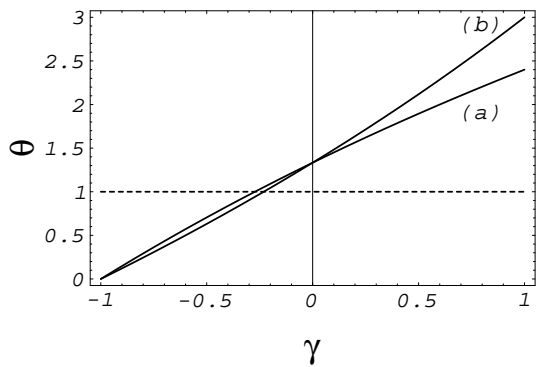}} 
\vspace{0.5cm}
\caption{(a) $\theta$ and (b) $\widetilde\theta$ in unit $\pi i$
as functions of $\gamma$ for $s_1=1/2$ and $s_2=1$.}
\label{f:Phase1}
\end{figure}
For example, 
in Fig.\ref{f:Phase1}, we explicitly show $\theta$
(see the curve (a) in the figure) as a function of $\gamma$ in
the case $s_1=1/2$ and $s_2=1$. We can see that at $\gamma\sim-0.273$,
$\theta$ becomes 1  and 
massless phase is expected.
It is interesting to compare this value with the numerical one 
$\gamma\sim-0.13$ in ref.\cite{TONE}.
This difference is not serious, as has been already discussed.
What we are interested in is mainly how many times
we expect massless phases in varying the bond-alternation parameter.
Indeed, the conclusion of Tonegawa et al\cite{TONE} is that there
occurs only one massless phase in the whole antiferromagnetic region 
of $\gamma$, which is consistent with our results.

As we have now two kind of spins, it is interesting to study an 
extreme case where the difference between the two spins is large.
Set, e.g,  $s_2\rightarrow\infty$, 
and we have $\theta\rightarrow 4\pi is_1$
(and $g\rightarrow0$) near $\gamma=1$. 
In this limit, therefore, we have $2s_1$ massless points
during the process $\gamma=-1\rightarrow 1$.
Namely, it turns out that smaller spin solely
 controls the number of the massless points.

Finally, we would like to compare the present result 
with the previous one (\ref{OldTGVB})
which was derived by neglecting the effects of the massive 
spin modes. It turns out  from (\ref{OldTGVB}) that
$\widetilde\theta=0\rightarrow2\pi i(s_1+s_2)$
for $\gamma=-1\rightarrow1$.
Therefore, in the case that $s_1$ and $s_2$ are the same type of spins 
(integers or half-integers) the model has $s_1+s_2$ massless
points, while it may have $s_1+s_2-1$ massless points in the case 
where the spins are of 
different type (the $\gamma=1$ point cannot be classified as a
massless point because the system is separated into disconnected 
dimers there).
In Fig.\ref{f:Phase1}, we present $\widetilde\theta$ for reference,
from which we see that behavior near $\gamma=1$ is different
from that of $\theta$. However,
 the value of $\gamma$ which gives $\widetilde\theta=1$ is
similar to that for $\theta$ in this case.
In general, $\theta\le\widetilde\theta$ at $\gamma=1$, where
equality holds only if $s_1=s_2$.
This discrepancy becomes more serious if the difference 
between the two spins $s_1$ and $s_2$ becomes larger.

Then the question is which formula is more reliable. 
\begin{table}[h] 
\begin{center}
\renewcommand{\arraystretch}{1.2}
\begin{tabular}{ll|lllll}
$\hfill s_1$& & 1/2 & 1 & 3/2 & 2 & 5/2\\ 
\hline
             &$v_{\rm sw}$  & 0.436 & 0.335 & 0.279 & 0.255 & 0.242 \\
$\gamma=0.9$ &$v$           & 0.436 & 0.336 & 0.279 & 0.255 & 0.242 \\
             &$\widetilde v$& 0.436 & 0.722 & 1.089 & 1.504 & 1.952 \\
\hline
             &$v_{\rm sw}$  & 0.436 & 0.614 & 0.748 & 0.859 & 0.955 \\
$\gamma=-0.9$&$v$           & 0.436 & 0.614 & 0.748 & 0.859 & 0.955 \\
             &$\widetilde v$& 0.436 & 0.591 & 0.689 & 0.768 & 0.837 \\
\end{tabular}
\end{center}
\caption{Spin-wave velocity for the Model B with bond-alternation. 
The spin $s_2$ is fixed as $s_2=1/2$, 
and other parameters are same as Table \ref{t:VelB}.}
\label{t:VelBbond}
\end{table}
In table \ref{t:VelBbond}, we compare the spin-wave velocities
calculated by means of the above two methods in case of 
the  finite bond-alternation,
corresponding to Figs.\ref{f:ModBbondP} and \ref{f:ModBbondM}.
It is clearly seen from this table as well as Table \ref{t:VelB}
that not only for $\gamma=0$ but also
for rather large $\gamma$, $v$ in (\ref{TheGVB})
exactly reproduces the spin-wave velocity $v_{\rm sw}$ 
calculated in the spin-wave analysis in the previous section.
On the contrary, the velocity $\tilde v$ 
considerably deviates from $v_{\rm sw}$.
Notice especially the difference between 
Figs.\ref{f:ModBbondP} and \ref{f:ModBbondM}.
These figures tell us that the hybridization  
between two spin modes in Fig.\ref{f:ModBbondP}
is much larger than that in Fig.\ref{f:ModBbondM}.
Therefore, if we naively neglect the massive mode,
it may cause more serious errors for the former case.
One can indeed see this fact in the above table: 
Namely, $\widetilde v$'s for $\gamma=0.9$ is worse in general than 
those for $\gamma=-0.9$. Based on these observations, we believe
that the present formulae eq.(\ref{TheGVB}),
which incorporate the effect of massive modes,
provide more reliable results.

To conclude, we may predict how many massless phases we encounter 
when we vary $\gamma$ from $-1$ to 1: Namely it 
is given by the number of odd-integers included 
in the region between 0 and $4s_1s_2(s_1+s_2)/(s_1^2+s_2^2)$.
It may be  an interesting issue 
to investigate this conjecture directly by numerical
calculations. 
\begin{figure}[h]
\epsfxsize=7cm 
\centerline{\epsfbox{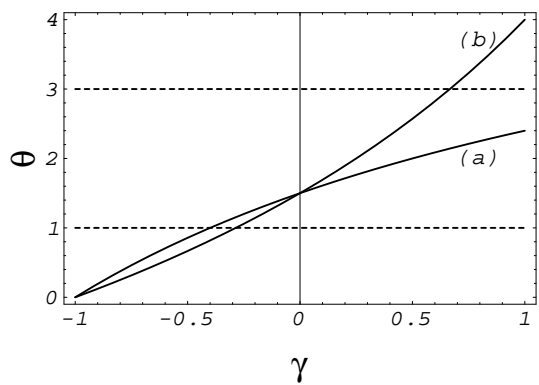}} 
\vspace{0.5cm}
\caption{(a) $\theta$ and (b) $\widetilde\theta$ in unit $\pi i$
as functions of $\gamma$ for $s_1=1/2$ and $s_2=3/2$.}
\label{f:Phase2}
\end{figure}
The simplest case is, for example, $s_1=1/2$ and
$s_2=3/2$. The present formula ($\theta$) 
predicts only one massless phase, 
while the previous formula ($\widetilde\theta$) 
predicts 2 massless phases, as can be seen from Fig.\ref{f:Phase2}.

\section{Summary}

We have investigated 
two kind of alternating spin chains, both of
which have singlet ground state.
They are composed of a periodic array of two kind of spins.
To analyze low-energy properties of these models, we have first 
studied spin-wave spectrum in detail.
Introducing quantum fluctuations semiclassically,
 we have converted the low-energy spin mode 
to the O(3) non-linear sigma model.
Analyzing the topological term in the sigma model, we have then 
clarified how the 
massless or massive excitations are developed reflecting
the topological nature of the alternating spin system.
Up to now, the alternating spin chains found experimentally 
have ferrimagnetic ground state.  We think that 
the systems similar to those proposed here with singlet ground state
could be fabricated experimentally, they would provide
a new paradigm of the quantum spin systems, for which
one may systematically observe the competition of the massive  and 
massless phases.

\acknowledgements
The authors would like to thank M. Chiba
for valuable discussions, and T. Tonegawa for 
fruitful discussions about their results prior to
publication. This work is partly
supported by the Grant-in-Aid from the Ministry of
Education, Science and Culture, Japan.

\appendix
\section{Formulae without introducing massive modes}

In this appendix, in comparison with the formulae 
given in the text, we write  down the expression without 
taking into account the effects of massive modes.
We here quote the results obtained in the previous
paper \cite{FukKaw}, in which  a preliminary 
treatment was given for low-energy properties of the model.
There, in order to map the model to
the non-linear sigma model, only
 a single $l$-field was introduced, which means 
 that the effects due to massive modes are 
neglected.  One can find 
the detail of the derivation is given in  the paper
\cite{FukKaw}. The formulae obtained are
\begin{eqnarray}
&&\widetilde\theta=2\pi is_1,
\nonumber\\
&&\widetilde g=\frac{2}{s_2}\frac{3}{\alpha+2},
\nonumber\\
&&\widetilde v=2Ja_0s_2\frac{2\alpha+1}{\alpha+1},
\label{OldTGVA}
\end{eqnarray}
for Model A, and 
\begin{eqnarray}
&&\widetilde\theta=2\pi is_{\rm eff}(1+\gamma_{\rm eff}),
\nonumber\\
&&\widetilde g=\frac{2}{s_{\rm eff}\sqrt{1-\gamma_{\rm eff}^2}},
\nonumber\\
&&\widetilde v=2Ja_0s_{\rm eff}\sqrt{1-\gamma_{\rm eff}^2},
\label{OldTGVB}
\end{eqnarray}
for Model B with $\gamma_1=\gamma_2\equiv\gamma$,
where
\begin{eqnarray}
&&s_{\rm eff}=\frac{s_1+s_2}{2},
\nonumber\\
&&\gamma_{\rm eff}=\frac{(s_1+s_2)^2\gamma-(s_1-s_2)^2}
{(s_1+s_2)^2-(s_1-s_2)^2\gamma}.
\end{eqnarray}


\end{document}